\documentstyle[aps,prb,epsfig]{revtex}
\begin{document}

\twocolumn[\hsize\textwidth\columnwidth\hsize\csname@twocolumnfalse\endcsname
\draft

\title{Theory of electron transport in normal metal/superconductor junctions}

\author{Xin-Zhong Yan$^{1,2}$, Hongwei Zhao,$^1$ and Chia-Ren Hu$^1$}

\address{$^1$Department of Physics, Texas A\&M University, College Station, TX 77843-4242 \\
$^2$Institute of Physics, Chinese Academy of Sciences, P. O. Box 603, Beijing 100080, China} 

\maketitle

\widetext
\begin{abstract} 
On the basis of the Keldysh method of non-equilibrium systems, we develop a theory of electron tunneling in normal-metal/superconductor junctions. By using the tunneling Hamiltonian model (being appropriate for the tight-binding systems), the tunneling current can be exactly obtained in terms of the equilibrium Green functions of the normal metal and the superconductor. We calculate the conductance of various junctions. The discrepancy between the present treatment and the well-known scheme by Blonder, Tinkham, and Klapwijk is found for some junctions of low interfacial potential barrier. 
\end{abstract}

\pacs{PACS numbers: 74.40.+r, 73.40.Gk}
\vfill
\narrowtext

\vskip2pc]

\section {Introduction}

One of the powerful methods detecting the quasiparticle states in a superconductor is to measure the conductance of a junction made up of a normal-metal and a superconductor (NS). There have been developed many theories describing the electron-tunneling phenomenon. In the case of the high interfacial potential-barrier limit, the linear-response theory is a well-known description.$^1$ But it is not valid for describing the electron transport in the low potential-barrier limit.

To calculate the conductance in more general cases, Blonder, Tinkham, and Klapwijk (BTK) have developed a theory by supposing that the system is in such a non-equilibrium state that only the incoming particles have equilibrium distributions.$^2$ This theory has been widely used for analyzing the tunneling phenomena in various NS junctions, and also has been extended for investigating electronic tunneling in Josephson junctions.$^3$

When a finite voltage is applied to a junction, the electron transport in the junction is a non-equilibrium process. We would like to consider the case when the current passing through the junction is a constant. The electron transport process is then a steady state. Such a non-equilibrium problem can be solved by the Keldysh approach.$^4$ In fact, this approach has been applied by a number of investigators for studying the tunneling in junctions of normal metals $^{5-6}$ and the electron transport under impurity scattering.$^7$

In this paper, we present a tunneling theory along this direction. We will start with a tunneling-Hamiltonian model defined in a square lattice. This model is appropriate for the tight-binding systems. The tunneling current can be exactly obtained in terms of the equilibrium Green functions of the normal metal and the superconductor. By so doing, all the effects of external voltage on the tunneling current can be rigorously taken into account. Moreover, it can be extended to study the tunneling in the point-contact junctions as in the scanning-tunneling microscope measurement. 

\section {Formalism}

We consider a junction consisting of a normal metal on the left side and a superconductor (SC) on the right side. In the Nambu representation, the tunneling Hamiltonian describing the electron-transport processes in the junction is given by
$$
H_T=\sum_{lr}(c_r^{\dagger }\hat T_{rl}c_{l}+c_l^{\dagger }\hat T_{lr}c_{r}) \eqno(1)
$$
where $c_{r}^{\dagger} = (c_{r\uparrow}^{\dagger}, c_{r\downarrow})$ is the field operator for particles in the right superconductor, and $c_{l}^{\dagger}$ is similarly defined for the left metal, $\hat T_{rl}=\hat T_{lr}^{\dagger} = t_0(|y_r-y_l|)\sigma_3$, and $y_r$ and $y_l$ are respectively the coordinates of the sites $r$ and $l$ along the interface. The $r$ and $l$ summations in Eq.(1) run over the edge (interface) sites on the two sides of the junction, respectively. The function $t_0(|y_r-y_l|)$ may be taken as real. For simplicity of description, we suppose that the lattice sites $\{r\}$ along the edge are equally spaced as the same as $\{l\}$. Suppose there is a voltage $V$ applied between the junction, the total Hamiltonian of the system is given by
$$H = H^0 + H_T \equiv H_l -eVN_l + H_r + H_T, \eqno(2)$$
where $H_l$ and $H_r$ are the intrinsic Hamiltonians of the left metal and the right superconductor, respectively, and $N_l$ is the total electron number of the left metal. We here adopt the tight-binding model for $H_r$ which contains a hopping term and an attraction term. For $H_l$, we keep only the hopping term.  

To define the tunneling-current operator, we first consider the charge operator for the right SC. Apart from a constant, it can be written as 
$$
Q=-e\sum_{r}c_r^{\dagger }\sigma_3c_{r}. \eqno(3)
$$
The operator of current through the junction from left to right is then obtained as
$$\hat I = i[H,Q] = ie \sum_{lr}(c_r^{\dagger }\sigma_3\hat T_{rl}c_{l}-c_l^{\dagger }\hat T_{lr}\sigma_3c_{r}).\eqno(4)$$

Now, let us choose the unperturbed state described by $H_0$ as our reference system. This reference system consists of the unperturbed normal metal and the SC on two side of the junction, each of them in its own equilibrium state. For the purpose of employing the grand canonical ensembles, we use $K_l = H_l -(\mu_l +eV)N_l$ and $K_r = H_r -\mu_rN_r$ to describe the normal metal and the Sc, respectively. Here, $\mu_l$ and $\mu_r$ are respectively the chemical potentials of the normal metal and the SC, and $N_r$ is the total number of electrons in the SC. At the steady state, we have $\mu_r = \mu_l+eV$ in order to maintain charge neutrality in the bulk of each side. To calculate the statistical average of a physical quantity, we need to write the related operators in the interaction picture. 
An operator of physical quantity, e.g., the current $\hat I(t)$, in the interaction picture at time $t$ is defined as, 
$$\hat I(t) =\exp(iH^0t)\hat I\exp(-iH^0t).$$
This operator can be further rewritten in terms of the field operators, 
$$\hat I(t) = -2e\,{\rm Im}\sum_{lr}c_r^{\dagger }(t)\sigma_3\hat T_{rl}(t)c_{l}(t),\eqno(5)$$
where $c^{\dagger}_r(t) = \exp(iK_rt)c^{\dagger}_r\exp(-iK_rt)$ (and a similar definition for $c_l(t)$), $\hat T_{rl}(t) = \hat T^{\dagger}_{lr}(t) = \hat T_{rl}\exp(ieVt\sigma_3)$. The form for 
$\hat I(t)$ as given by Eq. (5) is convenient for the statistical average over the grand canonical ensembles. Similarly, the tunneling Hamiltonian can be written as  
$$H_T(t) = \sum\limits_{lr}[c_r^{\dagger }(t)\hat T_{rl}(t)c_{l}(t)+c_l^{\dagger }(t)
\hat T_{lr}(t)c_{r}(t)].\eqno(6)$$

For applying the Keldysh method, it is convenient to define the field operator,
$$\phi^{\dagger}_r(t) = [c^{\dagger}_r(t_{+}), c^{\dagger}_r(t_{-})] \eqno(7)$$
where the subscripts + and - on time $t$ means the operators defined in the time branches ($-\infty, \infty$) and ($\infty, -\infty$), respectively. Accordingly, we define a perturbation Hamiltonian,
$$H_c(t) = \sum_{lr}[\phi_r^{\dagger }(t)T^c_{rl}(t)\phi_{l}(t)+\phi_l^{\dagger }(t)
T^c_{lr}(t)\phi_{r}(t)] \eqno(8)$$
where 
$$T^c_{rl}(t) = \left(\matrix{\hat T_{rl}(t) & 0 \cr 0 & -\hat T_{rl}(t)}
\right) \equiv
\tau_z\hat T_{rl}(t). \eqno(9)$$
The matrix $\tau_z$ is the third Pauli matrix defined in the space corresponding to the two time branches. To distinguish with that, we reserve $\sigma_3$ as the third Pauli matrix defined in the particle-hole space. The Green function is defined as
$$G_{ij}(t,t')=-i\langle {\cal T} [S_c\phi_i(t)\phi^{\dagger}_j(t')]\rangle$$
$$S_c = {\cal T} \exp[-i\int_{-\infty}^{\infty}dtH_c(t)]$$
where ${\cal T}$ is the Keldysh time-ordering operator. 

With the above definitions, the current under the statistical average can be expressed as
$$I = e \sum_{lr}{\rm Re\,Tr}\,\sigma_3\hat T_{rl}(t)G_{lr}(t,t),\eqno(10)$$
To calculate the current, we need to know the Green function $G_{lr}(t,t)$. It can be determined from the Dyson equations. 

Let $L$ and $R$ denote the Green functions (as $4\times 4$ matrices) for the left metal and the right SC, respectively (with the superscript 0 for the unperturbed ones). By assuming that the system is uniform along the direction parallel to the interface, we can then work in the momentum space. Here, the momentum is parallel to the interface. The Dyson equations are
$$G_k(t,t') = \int dt_1L^0_k(t,t_1) T^{c\dagger}_k(t_1)R_k(t_1,t') \eqno(11)$$
$$R_k(t,t') = R^0_k(t,t') + \int dt_1\int dt_2R^0_k(t,t_1)\Sigma_k(t_1,t_2)R_k(t_2,t') \eqno(12)$$
$$\Sigma_k(t_1,t_2) = T^c_k(t_1)L^0_k(t_1,t_2)T^{c\dagger}_k(t_2) \eqno(13)$$
where $T^c_k(t) = \tau_z\hat T_k\exp(i eVt \sigma_3)$, $\hat T_k = t_0(k)\sigma_3$, and the range of time integrals is from $-\infty$ to $\infty$. Note that the Green function $L^0_k(t_1,t_2) = L^0_k(t_1-t_2)$ consists of four diagonal matrices. The factors $\exp(i eVt_1 \sigma_3)$ and $\exp(i eVt_2 \sigma_3)$ commute with the matrix $L^0_k(t_1,t_2)$. The self energy $\Sigma_k(t_1,t_2)= \Sigma_k(t_1-t_2)$, and thereby the Green function $R_k(t,t') = R_k(t-t')$ are functions of time difference. We can therefore take the Fourier transformation of the Dyson equations. In the frequency space, these equations have the usual forms except
$$\Sigma_k(\omega) = T^c_k(0)L^0_k(\omega+eV\sigma_3)T^{c\dagger}_k(0). \eqno(14)$$
With the help of the Dyson equations, we can write the factor $\hat T_{rl}(t)G_{lr}(t,t)$ in the expression of $I$ as
$$\hat T_{rl}(t)G_{lr}(t,t) = \int_{-\infty}^{\infty}{d\omega\over 2\pi}\tau_z\Sigma_k(\omega)R_k(\omega).\eqno(15)$$
Inserting Eq. (15) into Eq. (10) and taking the trace of time-branch space, we have
$$I=e\sum_{k}\int^{\infty}_{-\infty}{d\omega\over 2\pi}t_0^2{\rm Re Tr}\sigma_3M_{+}(L_fR^0_{-}+L^0_{+}R_f)M_{-}
\eqno(16)$$
with
$$M_{\pm}=[1-t_0^2L^0_{\pm}R^0_{\pm}]^{-1},$$
$$L_f = \tanh[(\omega+eV\sigma_3)/2k_BT](L^0_{+}-L^0_{-}),$$ 
$$R_f = \tanh(\omega/2k_BT)(R^0_{+}-R^0_{-}),$$
$$L^0_{+} = L^{0\dagger}_{-} = L^0(k,\omega+eV\sigma_3+i0),$$
$$R^0_{+} = R^{0\dagger}_{-} = R^0(k,\omega+i0).$$
Here $L^0_{+}$ and $R^0_{+}$ ($L^0_{-}$ and $R^0_{-}$) are the retarded (advanced) Green functions (as $2\times 2$ matrices in the Nambu space) of equilibrium state, $L_f$ and $R_f$ are the Keldysh functions, $t^2_0 = |t_0(k)|^2$, $k_B$ is the Boltzmann constant, and $T$ is the temperature of the system. By noting the relationships $R_{+}(k,-\omega) = -\sigma_2 R_{-}(k,\omega)\sigma_2$, $L_{+}(k,-\omega+eV\sigma_3) = -\sigma_2 L_{-}(k,\omega+eV\sigma_3)\sigma_2$, it is enough to only take the frequency integral in Eq. (16) in the range $(0, \infty)$. The front factors in the Keldysh functions take part of the roles of quasiparticle distribution functions. The additional term $-eV\sigma_3$ reflects the chemical potential shifts of the quasiparticles in the left metal. 

\section {Green's functions of the equilibrium state}

To calculate the tunneling current $I$, we need to know the Green functions $L^0$ and $R^0$.
If we know the wave functions $\psi_n$ and energies $E_n$ of the quasiparticles, e.g., for the SC, we can obtain $R^0$ by
$$R^0(k,\omega)= \sum_n {\psi_n\psi^{\dagger}_n\over \omega-E_n},\eqno(17)$$
where $\psi_n$ takes the edge value. Since we have taken the Fourier transformation for the dependence on the coordinates parallel to the interface, the wave function $\psi_n(j)$ depends on the $x$-coordinates (normal to the edge) of lattice sites, $j = \{1,2,\cdots\}$; the edge value is $\psi_n(1)$. 

For illustration, we here consider a $d$-wave SC and suppose that the order parameter is constant everywhere. The wave functions can be determined analytically by the BdG equation. As an example, we consider the tight-binding model defined in a semi-infinite square lattice with a \{11\} edge. The BdG equation reads $^8$
$$\sum_{j}H_{ij}\psi_n(j) =E_n \psi_n(i), \eqno(18)$$
where $H_{jj} = -\mu\sigma_3$, $H_{j,j-1} = -2t\cos k\sigma_3 - i2\Delta\sin k\sigma_1$ for $j \ge 2$, $H_{j,j+1} = -2t\cos k\sigma_3 + i2\Delta\sin k\sigma_1$, otherwise $H_{ij} = 0$, $t$ is the hopping energy of electrons between nearest-neighbor sites, and $\Delta$ is the order parameter. Here, we have used the unit $\sqrt{2}/a$ (with $a$ the lattice constant) for the momentum $k$, and $k$ is confined to a Brillouin zone $(-\pi/2, \pi/2)$. There are two kinds of solutions to Eq. (18): The continuum states and the surface bound states.

The continuum states are generally degenerate. To distinguish them, we can consider each eigen wave function contains a unique incoming wave component or a unique outgoing wave component. We then characterize the wave function by the incoming wave number $q_{\mu}$ or the outgoing wave number $q_{\alpha}$. For example, the wave function and energy of state $q_{\mu}$ can be written as
$$\psi_{k,\mu}(j) = [\psi^0_{k,\mu}(j)-\sum_{\alpha}a_{\mu\alpha}\psi^0_{k,\alpha}(j)]/\sqrt{2}, \eqno(19)$$
$$E_{k,\mu} = \pm E(q_{\mu},k) = \pm\sqrt{e^2(q_{\mu},k)+\Delta^2(q_{\mu},k)},\eqno(20)$$
where $\psi^0$'s are the plane-wave solution to the infinite system, $e(q,k) = -4t\cos q \cos k -\mu$ (with $\mu$ the chemical potential), $\Delta(q,k) = -4\Delta\sin q \sin k$. The coefficients $a_{\mu\alpha}$ are determined by the boundary condition at $j = 1$. The summation over $\alpha$ in eq. (19) runs over all the outgoing components with $E(q_{\alpha},k)= E(q_{\mu},k)$. It is worth noticing that sometimes we may have complex $q_{\alpha}$'s, the summation then should be taken at those $q_{\alpha}$'s corresponding to decaying waves. 

The number of the bound states is determined by the Levinson theorem.$^9$ Under the assumption that the order parameter is constant, we only have the state with $E_n = 0$ for each $|k| \le k_m$ ($k_m$ is very close to the Fermi wave number).$^{8,10}$ For $E_n = 0$, it can be shown that the two components $u_k(j)$ and $v_k(j)$ satisfy the relation, 
$$v_k(j) = i\lambda u_k(j), ~~\lambda = \pm 1. \eqno(21)$$
Suppose $u_k(j) = z^j$ with $z$ ($|z| < 1$) a complex quantity for the general solution. Corresponding to $z$, we have a complex number $q = -i\log(z)$. The equation $E(q,k) = 0$ determining the eigenvalue reduces to
$$t(z+z^{-1})\cos k + \lambda(z-z^{-1})\Delta\sin k+ \mu/2 = 0. \eqno(22)$$
The solutions to Eq. (22) are
$$z_{\pm} = [-\mu\pm\sqrt{\mu^2 -(c_1^2-c_2^2)}]/(c_1+\lambda c_2), \eqno(23)$$
where $c_1 = 4t\cos k$ and $c_2 = 4\Delta\sin k$.
Note $z_{+}z_{-} = (c_1 - \lambda c_2)/(c_1 + \lambda c_2)$, therefore $\lambda = {\rm sgn}(k)$ whereby $|z_{+}z_{-}| < 1$. The wave function is given by
$$u_k(j) = (z_{+}^j-z_{-}^j)/N_k, \eqno(24)$$
with $N_k^2 = 2[(1-|z_{+}|^2)^{-1}+(1-|z_{-}|^2)^{-1}-2{\rm Re}(1-z_{+}^{*}z_{-})^{-1}]$ the normalization constant. This wave function satisfies the boundary conditions at $j =1$ and $j \to \infty$ provided $|z_{\pm}|< 1$. If $\mu^2 < (c_1^2-c_2^2)$, then $z_{+}$ and $z_{-}$ are complex conjugates of each other, and $|z_{\pm}| < 1$. On the other hand, if $\mu^2 > (c_1^2-c_2^2)$, both of them are real. In this case, there may be no bound state unless both $|z_{\pm}| < 1$.

With the knowledge of the wave functions, the Green function $R^0$ can be calculated by Eq. (17). As for $L^0$ of the normal metal, it contains only the continuum states. The wave functions can be obtained immediately from Eq. (18) by setting $\Delta = 0$. The resulted Green function is given by 
$$L^0(k,\omega) = {2\over \pi}\int_{0}^{\pi}dq{\sin^2q\over\omega-e(q,k)\sigma_3}.\eqno(25)$$

\section {Comparison with the BTK theory}

Obviously, the present treatment is a non-perturbative theory. It takes into account all the effects of the voltage within the model. At this point, it is instructive to compare our theory with the BTK theory. In the BTK model, only the incoming particles in each side of the junction are described by the equilibrium distributions with the chemical potential shift of the left metal due to the external voltage. But, the outgoing particles are not described by the equilibrium distributions. The quasiparticle states in the whole system are determined by the Bogoliubov-de Gennes equation that is independent of the external voltage. The tunneling current is calculated as the result of the current by the incident particles from the left metal minus that from the right SC. In contrast, by the present consideration, the particle distributions are referred to the reference system. Since in the interaction picture, the tunneling Hamiltonian depends on time, there cannot be quasiparticle states for the whole system. Each state in both sides of the junction has its lifetime because of the non-equilibrium process between the interface. From the Green function, the lifetime of a quasiparticle is determined by the inverse of the imaginary part of the self-energy. In this approach, the transport process is treated by the equivalent of time-dependent perturbation theory to all orders, which leads to lifetimes. The electron transport is the process of quasiparticles decaying. On the other hand, in the BTK model, the transport process is treated by the time-independent perturbation theory to all orders, which determines the quasiparticle states in the whole system, with infinitive lifetimes for the continuum states. Therefore, the mechanisms of electron transport through the junction by the two theories are very different.

For numerical comparison, we need to present the BTK scheme in the lattice model. The basic work in the scheme is to solve the BdG equation for the wave functions of quasiparticles in the whole system. An eigen wave function characterized by an incoming wave in the left metal can be written as the incoming wave plus all the reflected waves (including the Andreev and the ordinary reflections), with the transmitted waves in the right SC including all the outgoing waves. One needs only then consider the boundary condition at the interface barrier. By denoting the wave functions in the left and right sides respectively by $\psi_l(j)$ with $j = \{-1, -2, \cdots\}$ and $\psi_r(j)$ with $j = \{1, 2, \cdots\}$, the BdG equation at the interface barrier reads
$$H_{-1,-2}\psi_l(-2)+H_{-1,-1}\psi_l(-1)+\hat T^{\dagger}_k\psi_r(1) = E \psi_l(-1), \eqno(26a)$$
$$\hat T_k\psi_l(-1)+H_{1,1}\psi_r(1)+H_{1,2}\psi_r(2) = E\psi_r(1). \eqno(26b)$$
Eqs. (26a,b) are nothing but the boundary conditions. With the wave functions, one can immediately calculate the tunneling current according to the BTK theory.

To see the difference between the present and the BTK theories, we have carried out the numerical calculations of the tunneling conductance 
$$G = {dI\over dV} \eqno(27)$$
for normal-metal/$d$-wave superconductor junctions with $\{110\}$ and $\{100\}$ interface at various barrier strengths. For presentation, we normalize $G$ by $Ne^2/\pi$ ($\hbar=1$) with $N$ the total number of the lattice sites on one side of the interface. The basic parameters for the SC are, $t = 176 meV$, hole concentration $\delta = 0.15$, attractive potential between the nearest-neighbor sites $v = 124 meV$. The transition temperature $T_c$ and the order parameter $\Delta_0$ are obtained as $T_c = 90 K$, and $\Delta_0 \equiv 4\Delta|_{T=0} = 16.7 meV$, respectively. As being stated before, the Hamiltonian of the left metal contains only the hopping term. We assume that the hopping energies of both sides of the junction are the same. For simplicity, we choose tunneling matrix element as $t_0(k) = t_0$. 

The numerical result for the normalized conductance as function of $V$ for an NS ($d$-wave) junction with \{100\} interface at $T = 0$ is shown in Fig. 1. The tunneling parameter $t_0/t = 0.5$ is used. Though the interfacial potential barrier at this parameter is not very high, the agreement between BTK and the present theories is very good. A small $t_0$ means a high interfacial potential barrier. At the high potential barrier limit, both theories reproduce the linear response result [1]. However, at $t_0/t = 1$ corresponding to a weak barrier, the discrepancy is clear as shown in Fig. 2. At weak barrier and small voltage $|eV| \le \Delta_0$, the Andreev reflection is the predominant contribution to the conductance in the BTK theory. Under the present assumption, however, the transport is due to the decay of quasiparticles in both sides. Such a decaying process is more complex than the BTK picture. The difference between the two theories at small $|eV|$ is mainly due to the different treatment of the tunneling Hamiltonian (i.e., time-dependent vs time-independent perturbation theory). The voltage effect in $L^0$ is important only at large $|eV|$, because the relevant dimensionless parameter is the ratio $eV/E_F$ (with $E_F$ the Fermi energy of the left metal). The voltage effect is more evident at $V < 0$ than at $V > 0$, because more precisely the parameter is actually $|eV/(E_F + eV)|$. At negative voltage, the chemical potential of the left metal shifts upward, resulting in electrons right below the Fermi surface within the energy range $(E_F+eV, E_F)$ transferring into the right SC. At positive voltage, the states in the energy range $(E_F, E_F+eV)$ in the left metal are available for the electrons in the right SC to transfer in.  

In Fig. 3, we show the results for the junctions with \{110\} interface at $t_0/t = 1$. In this case, the results by both theories are in excellent agreement. The agreement is even better at smaller $t_0$. At $|eV| < \Delta_0$, the conductance $G$ is given by a broadened zero-bias peak. Actually, there are zero-energy bound states in the right SC near the interface, with lifetime due to tunneling. The tunneling current is predominantly conducted by these states. The width of the broadening is mainly determined by the tunneling parameter $t_0$ rather than by the external voltage. Because of the existence of these states, the particle transmission through the junction for $|eV| < \Delta_0$ is a resonant process. These resonance states exist in the BTK model as well. At least at $eV = 0$, both theories produce the same resonance states with the same energy broadening. Therefore, we can understand the excellent agreement near $eV = 0$. 

The discrepancy between the two theories is even more clear for the normal-metal/conventional-superconductor junctions. Fig. 4 shows the result for an NS ($s$-wave) junction with $\{100\}$ interface at $T = 0$ and $t_0/t = 1$. The parameters for the SC are, the chemical potential $\mu = -0.3t$, the on-site pairing parameter $\Delta_0 = 0.02t$. The conductance predicted by the present theory is only about $78\%$ of that of BTK for $|eV|/\Delta_0 \le 1$ where the conductance is almost a constant. Qualitatively, the electron transport in this junction is similar as that in the NS ($d$-wave) junction with \{100\} interface. The explanation for Fig.2 applies here.

\section{An approximation scheme}

When $eV/E_F \ll 1$, the dependence of $L^0$ on the external voltage is very weak. We can then drop $eV$ in $L^0$. By this approximation, the conductance $G$ is given by
$$G =-2e^2\sum_{k}\int^{\infty}_{-\infty}{d\omega\over 2\pi}t_0^2{\rm Tr Im}(R^0_{-}M_{-}\sigma_3M_{+})\sigma_3{\rm Im}L^{+}g
\eqno(28)$$
where $g = \cosh^{-2}[(\omega+\sigma_3eV)/2k_BT]/2k_BT$ is only fact which depends on $eV$. In Fig. 4, the result by Eq. (28) is also plotted. At small voltage, the approximation is in very good agreement with our main theory. However, at large voltage, the approximation reproduces the BTK result. This clearly shows that the discrepancy between our main theory and the BTK theory at small voltage is not due to the voltage effect in $L^0$. In the case of $eV/E_F \ll 1$, Eq. (28) is a simple but good scheme for calculation of the conductance.

\section{Summary}

In summary, on the basis of the Keldysh approach, we have developed a theory of electron transport in normal metal/superconductor junctions to all orders in the applied voltage and the barrier strength. In the present scheme, the tunneling current is given in terms of renormalized Green functions of a steady state. It can give a reliable description of the electron tunneling, including the ballistic transport in NS junctions. We have calculated the tunneling conductance for various NS junctions using the present formalism and have compared it with the BTK theory. In most cases, both theories agree with each other. However, for some junctions of low barrier strength, the discrepancy between the two theories can be sizable.

\acknowledgments

This work is supported by the Texas Higher Education Coordinating Board under the grant No. 1997-010366-029, and by the Texas Center for Superconductivity at the University of Houston.

\vskip 25mm

\begin{figure}
\begin{center}
\epsfig{file=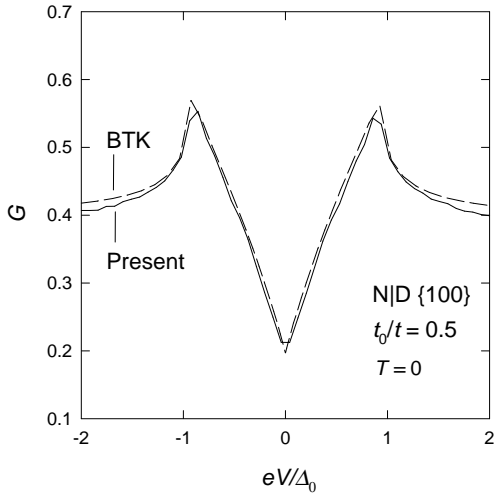,height=6.5cm, width=6.5cm}
\vskip 5mm
\caption
{Conductance $G$ as a function of the Voltage $V$ for an NS (d-wave) junction with \{100\} interface at $T =0$ and $t_0/t = 0.5$. The present calculation (solid line) is compared with the BTK result (dashed line).}
\end{center} 
\end{figure} 
\vskip 20mm
 
\begin{figure}
\begin{center}
\epsfig{file=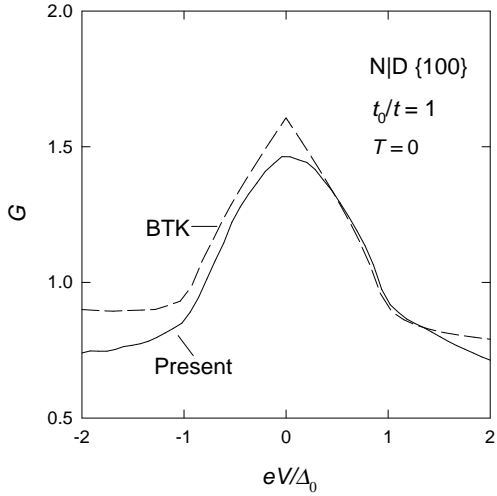, height=6.5cm, width=6.5cm}
\vskip 5mm
\caption{The same as Fig.1 but at $t_0/t = 1$.}
\end{center} 
\end{figure} 
\vskip 20mm
 
\begin{figure}
\begin{center}
\epsfig{file=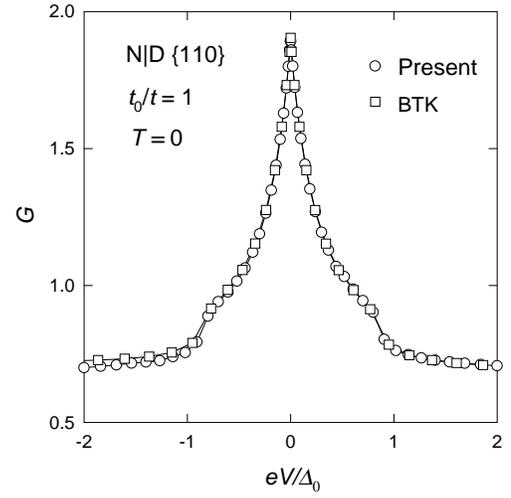,height=6.5cm, width=6.5cm}
\vskip 5mm
\caption
{Conductance $G$ as a function of the Voltage $V$ for an NS (d-wave) junction with \{110\} interface at $T =0$ and $t_0/t = 0.5$. The present calculation (circles) is compared with the BTK result (squares).}
\end{center} 
\end{figure}
\vskip 20mm
 
\begin{figure}
\begin{center}
\epsfig{file=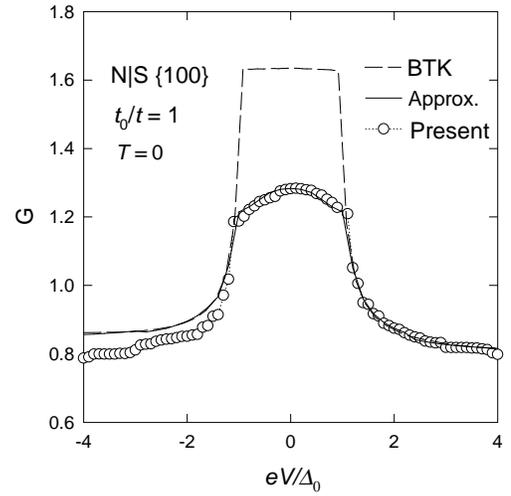,height=6.5cm, width=6.5cm}
\vskip 5mm
\caption
{Conductance $G$ as a function of the Voltage $V$ for an NS (conventional SC) junction with \{100\} interface at $T =0$ and $t_0/t = 1$. The present calculation (dotted line with circles) is compared with the BTK (dashed line), and the approximated (solid line) results.}
\end{center} 
\end{figure} 

\end{document}